\documentclass[aps,rmp,showkeys,nolinenumbers,amsmath,amssymb,longbibliography,a4paper,11pt,floatfix]{revtex4-2}

\pdfoutput=1

\usepackage{graphicx}
\usepackage{subfig}
\usepackage[font=small]{caption}
\usepackage[separate-uncertainty,detect-all,range-phrase=\ and\ ]{siunitx}
\usepackage[utf8]{inputenc}
\usepackage[T1]{fontenc}
\usepackage{bm}
\usepackage[english]{isodate}
\cleanlookdateon
\renewcommand{\BibitemShut}[1]{}

\usepackage{breakurl}
\usepackage{hyperref}

\begin{document}

%\title{Measurements of droplets from singing and some other activities}
\title{Measurements of droplets from singing, laughing, reciting poetry, and playing wind instruments}
\thanks{Originally, this manuscript has been published as a project memorandum, and can be downloaded here: \href{https://hdl.handle.net/11250/2888270}{https://hdl.handle.net/11250/2888270}. The only two unessential differences between the original and this version are (1) the slightly adapted title, and (2) an added sentence on supplemental material including the reference.}
\thanks{Supplementary material: \citet{Dunkeretal2020data}}
%thanks{\url{https://www.sintef.no/siste-nytt/2020/sprer-korsang-mer-draper-enn-latter-ny-sintef-studie-gir-svaret/}}
\date{\printdate{07.09.2020}}
\author{Tim Dunker}
\email{tim.dunker@sintef.no}

\author{Jon Tschudi}
\email{jon.tschudi@sintef.no}

\author{Marion O'Farrell}
\email{marion.ofarrell@sintef.no}

\affiliation{Applied Optics, SINTEF Digital, Forskningsveien 1, 0373 Oslo, Norway}

\begin{abstract}
We present initial results from measurements of exhaled droplets by two female singers during singing, speaking, laughing, and recitation of poetry. We also conducted measurements with a flute, a clarinet, a tuba, and only the tuba's mouthpiece. This small project is motivated by the current distancing restrictions for choirs, musicians, and actors. To be able to image and track droplets, we have developed and built a portable measurement set--up. We have detected droplets with diameters of approximately \SI{50}{\micro\metre} and above. We found that for a single subject, the largest amount of droplets are produced during singing, but that laughing can produce a comparable number of droplets. Speaking and recitation of poetry do not produce as many droplets. We repeated exercises at varying distances and found that the number of detected droplets decreased rapidly with increasing distance. However, the relative height difference between people, e.g. in a stage setting, must be considered when determining how far a droplet of a given size and velocity can reach. Most droplets we detected follow a ballistic trajectory, and hit the ground after a distance of approximately \SI{1}{\metre}. In case of the musical instruments, we did not detect droplets exiting a tuba and only few single droplets from a flute and a clarinet. A clarinet usually points downwards, such that droplets follow a downward trajectory and hit the ground after a short distance. The situation was different when the tubist used only the mouthpiece (that is, detached from the tuba) during warm--up while partially blocking the opening with a finger: in this case, many droplets were ejected from the mouthpiece. These followed ballistic trajectories at different speeds, with some droplets exiting the measurement volume. Two important limitations of this study are that (i) the sample size is very small, such that a statistical analysis is beyond our scope; and (ii) that we cannot draw conclusions about a possible risk of infection with COVID--19 when performing any of the said activities.

% Call 'showkeys' to print the keywords:
\end{abstract}

% Call 'preprint' option to the class to show this on the first page. It moves the abstract to the second page.
%\preprint{Report number}

\maketitle

\section*{Disclaimer}
The purpose of this small project was building a set--up and demonstrating that the generation of droplets from a limited number of subjects, during different activities, could be measured and compared. The set--up was designed and built so that it could be used in different scenarios. However, for these initial tests, all measurements were performed in an ordinary room at SINTEF.

Due to limited time and budget, measurements were performed on only two subjects. We must assume that different people generate different amounts of droplets. Therefore, a complete study would require a larger number of subjects, measured in different situations/environments. Our observations must, therefore, be treated as preliminary, and the conclusions are subject to change upon expansion of the project---for example, with a more in--depth analysis of a larger, more statistically representative number of measurements.

We only measure droplets that are larger than a certain diameter, and thus have enough mass causing them to fall to the ground after a limited distance. The smallest droplets, often referred to as ``aerosols'', that can be airborne for an extended time are not measured in this experiment. It is, however, assumed by the WHO that larger droplets and close contact is the most common way of getting infected by SARS--CoV--2\footnote{\href{https://www.who.int/news-room/commentaries/detail/transmission-of-sars-cov-2-implications-for-infection-prevention-precautions}{https://www.who.int/news-room/commentaries/detail/transmission-of-sars-cov-2-implications-for-infection-prevention-precautions}}.

The researchers involved in this project are neither certified nor qualified to make any conclusions related to the risk of infection based on these measurements.

It would have been preferable to SINTEF to publish the findings after the preliminary results were verified on a larger subject set, but due the acute situation and public concerns related to COVID--19, Norges Korforbund and the co--sponsors of the project wish to share these preliminary results. We are confident that that this work can make a solid basis for possible future investigations.

\clearpage

\section*{Executive summary}
We present initial results from a limited number of experiments. The main results focus on data measured of a professional and an amateur singer performing different activities such as singing, warm--up exercises, speaking, laughing, and poetry recitation. Both singers are female. We also did single recordings of a tubist, flutist, and clarinettist, each playing their respective instruments. Due to the limited number of subjects, this cannot be defined as a statistical study. We do not draw conclusions about a possible risk of infection with, e.g., COVID--19 based on our experiments. All measurements were conducted without a face mask. We detected and observed the behaviour of droplets with a diameter of approximately \SI{50}{\micro\metre} and above.

The results of this initial study show that singing produces more droplets relative to speaking or poetry recitation. The number of droplets varies with how the sound is formed---large tongue movements, pursing of lips and rapid opening and closing of the mouth produced most droplets in our experiments. Laughing can also produce droplet numbers comparable to singing. We also observe that, due to the downward trajectory of the droplets, it is just as important to consider the relative height between people as it is to consider the distance between two people, for example a person standing beside a sitting person, or people singing from different heights on a choir riser.

We did not detect a considerable number of droplets from the tip of either the flute or clarinet. We did not detect any droplets from a tuba. When only the mouthpiece of a tuba is used during warm--up and the opening was partially blocked by a finger, due to the way it was held, several droplets were ejected that followed ballistic trajectories.

The work presented here is indicative. However, it has the potential be more conclusive, if a more thorough study is conducted, with a larger subject set. The set--up is also portable, so that it can be used to test people in different locations and scenarios. This field work could also be part of the follow--on study.

\clearpage

\section{Introduction}
During the COVID--19 pandemic, several restrictions regarding distance to other human beings have been imposed by many governments. As of \printdate{31.08.2020}, the current Norwegian recommendations are that choir members and musicians keep a distance of \SI{1.5}{\metre} to the side and \SI{2}{\metre} to the front. Choirs, orchestras, and actors have had to limit or cancel events because of such restrictions. There have also been several media reports of COVID--19 infection cases related to choir gatherings, for example, in Amsterdam\footnote{\href{https://www.theguardian.com/world/2020/may/17/did-singing-together-spread-coronavirus-to-four-choirs}{https://www.theguardian.com/world/2020/may/17/did-singing-together-spread-coronavirus-to-four-choirs}}\footnote{\href{https://www.latimes.com/world-nation/story/2020-06-01/coronavirus-choir-singing-cdc-warning}{https://www.latimes.com/world-nation/story/2020-06-01/coronavirus-choir-singing-cdc-warning}}, on Vr\aa ng\"o\footnote{\href{https://www.svt.se/nyheter/lokalt/vast/coronautbrott-pa-vrango-efter-korsammankomst}{https://www.svt.se/nyheter/lokalt/vast/coronautbrott-pa-vrango-efter-korsammankomst}}\footnote{\href{https://www.svt.se/nyheter/lokalt/vast/coronautbrottet-mindre-an-befarat-pa-vrango}{https://www.svt.se/nyheter/lokalt/vast/coronautbrottet-mindre-an-befarat-pa-vrango}} off the coast of Gothenburg, and in Seattle\footnote{\href{https://www.seattletimes.com/seattle-news/health/mount-vernon-choir-outbreak-was-superspreader-event-says-cdc-report-on-how-easily-virus-spreads/}{https://www.seattletimes.com/seattle-news/health/mount-vernon-choir-outbreak-was-superspreader-event-says-cdc-report-on-how-easily-virus-spreads/}}\footnote{\href{https://www.cnn.com/2020/05/13/us/coronavirus-washington-choir-outbreak-trnd/index.html}{https://www.cnn.com/2020/05/13/us/coronavirus-washington-choir-outbreak-trnd/index.html}}.

The current consensus among epidemiological authorities is that infections through droplets is the predominant means of infection with COVID--19 (Ernst--Kristian R\o dland, pers. comm., 2020). But how far do droplets spread when you are singing, talking, or playing an instrument? Are there vocal activities and exercises that produce more droplets or farther--reaching droplets than others? The goal of this project is to answer these questions.

While several groups have optically measured particles that originate from speaking \citep{Anfinrudetal2020,Bahletal2020a,Stadnytskyietal2020}, coughing and sneezing \citep{Bahletal2020b,Bourouibaetal2014,Hanetal2013}, no articles detailing the direct measurement of droplets during singing and playing musical instruments were sourced during the work detailed in this memo.

We developed a portable set--up where particles can be measured and tracked in different environments. Tracking is possible due to the continuous illumination of a measurement volume (approximately $\SI{20}{\centi\metre}\times \SI{20}{\centi\metre} \times \SI{10}{\centi\metre}$) by a light--emitting diode (LED), while full high--definition videos are taken at a frame rate of \SI{100}{\hertz}. We recorded videos of a professional opera singer performing songs, reciting poetry, speaking, and laughing; musicians playing flute and tuba (specifically the mouthpiece); and an amateur singer doing warm--up exercises at different distances from the measurement volume. We compared the various activities, and the distance at which the generated particles could no longer be detected.

\subsection{Scope and limitations of this initial study}
The original goal of the study was limited to showing that it is possible to detect particles from different vocal and musical activities, possibly at varying distance. We did not have multiple subjects performing the identical activities for the purpose of statistical representation and comparison. This is a significant limitation of the work. It is difficult to quantify the results. Therefore, the scope of this initial study is limited to comparing different activities to each other, such as singing, speaking, or reciting a poem. Additionally, a tuba player performed warm--up exercises using only the mouthpiece. Initial results from these measurements are shown in this memorandum.

From the measurements and results presented here, it is not possible to draw conclusions about a risk of infection with COVID--19. This is beyond the scope of this study.

\section{Measurement principle and data analysis}
The measurement principle is based on the forward--scattering of visible light by particles. That is, we do not detect or sample the particles directly, but the light scattered by them. The experiments were conducted in two different rooms. The experiments with the professional opera singer were conducted in a room with considerably less dust.

\begin{figure}[!t]
\centering
\subfloat[Sketch of the optical set--up with a LED, two lenses, and a digital camera, as seen from above. The camera is placed such that the direct LED beam is not entering the camera lens. The aperture was selected such that only particles within the marked region were in focus, and dust on the lenses and particles outside the volume of interest were blurred. ]{\label{fig:1a}\includegraphics[width=0.5\textwidth]{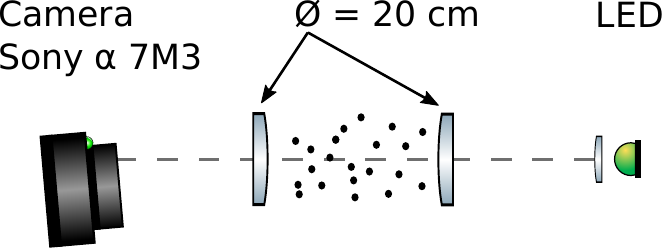}}\\
\vfill
\subfloat[Raw image, frame 6735 of time series labelled ``Singing'' in Fig. \ref{fig:3}. Droplets are somewhat visible in this image.]{\label{fig:1b}\includegraphics[width=0.5\textwidth]{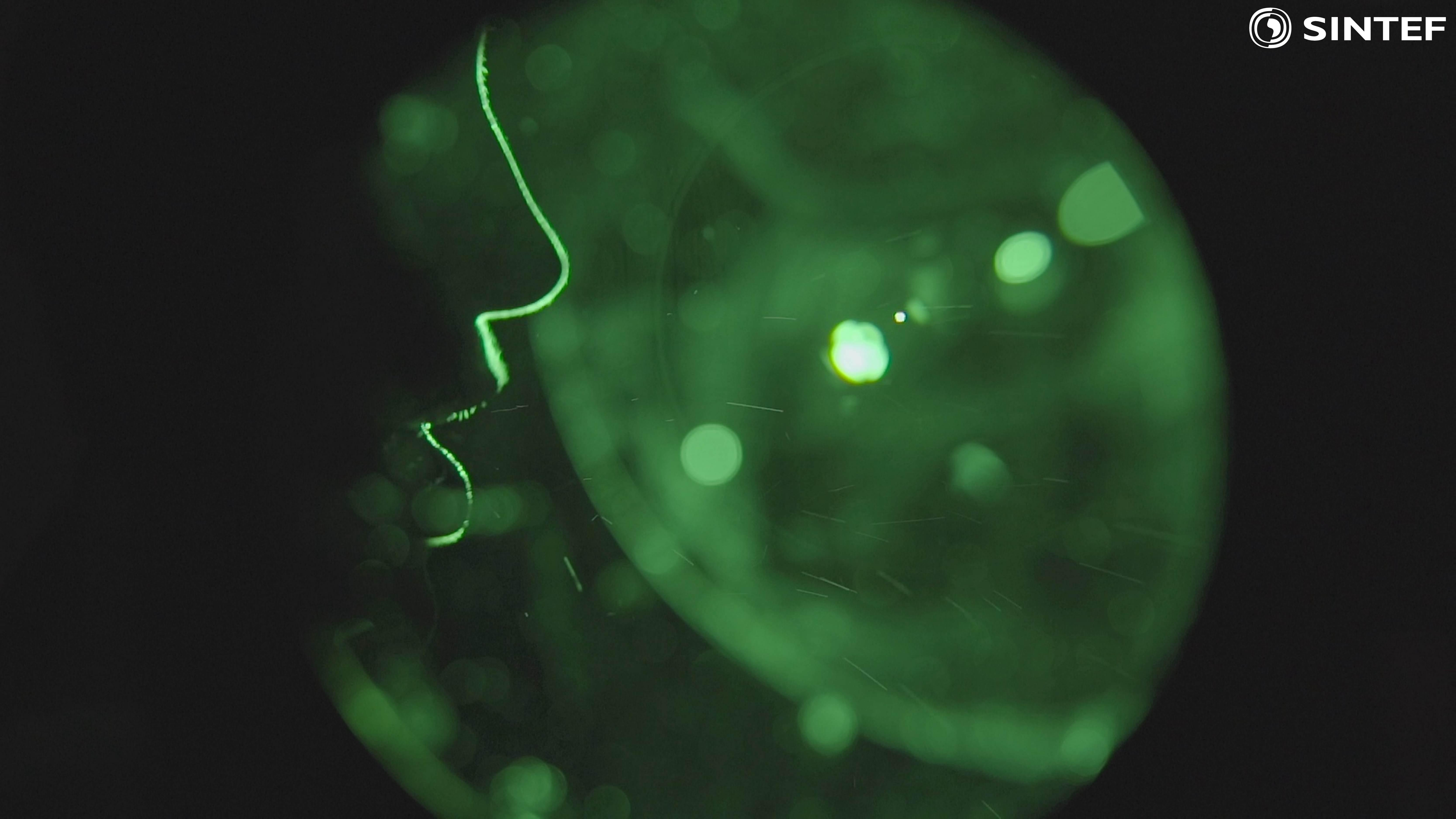}}\\
\vfill
\subfloat[Background--subtracted image with overlay of detected particles.]{\label{fig:1c}\includegraphics[width=0.54\textwidth]{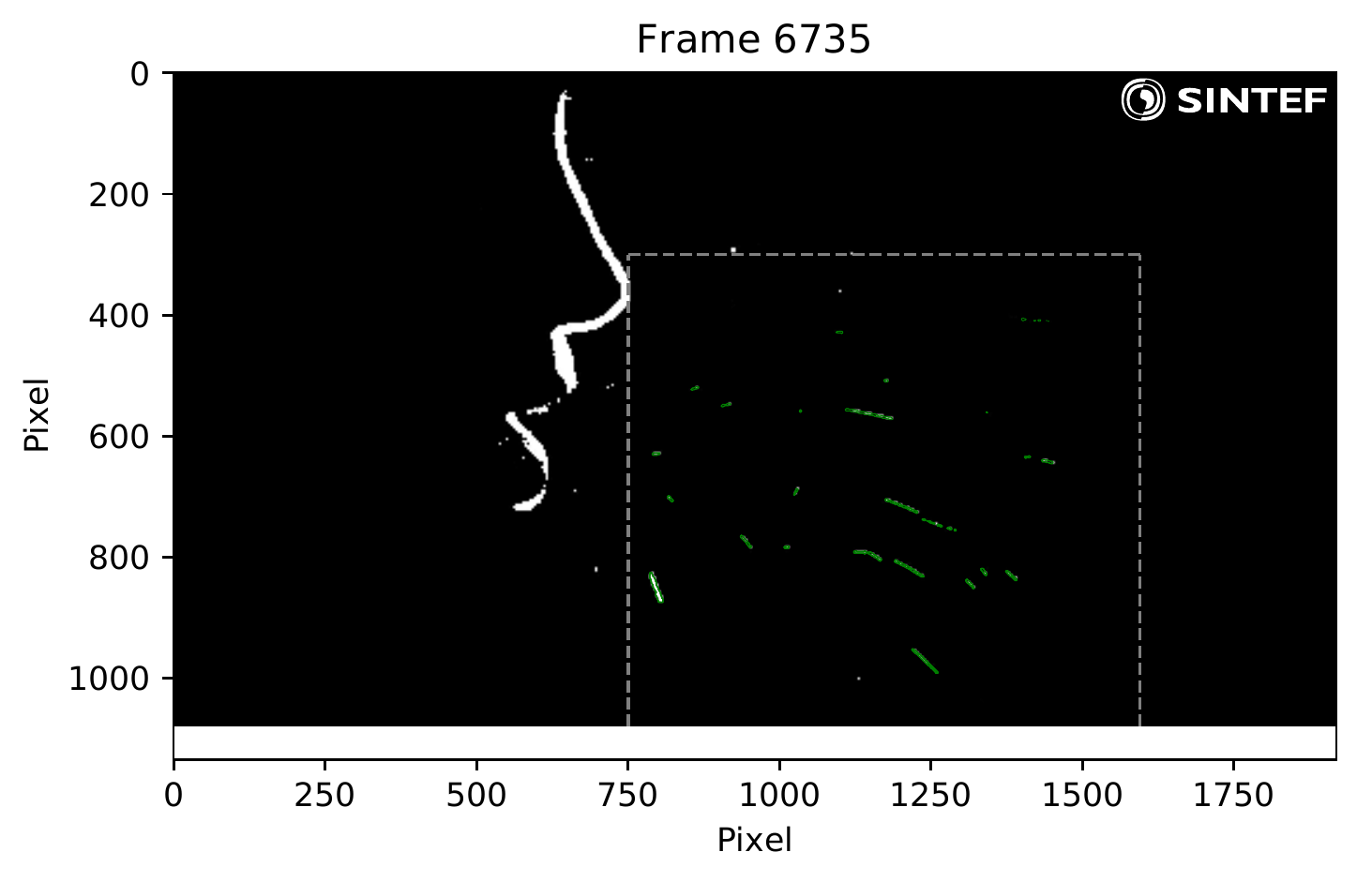}}
\caption{The measurement setup is shown in panel \ref{fig:1a}. Panel \ref{fig:1b} shows an example of a raw image and panel \ref{fig:1c} shows a composite of the background--subtracted image of \ref{fig:1b}, where the detected particles are marked in green. The dashed lines are the boundaries of the region where particles are detected. This frame is also part of the composite image shown in Fig. \ref{fig:4}, albeit with different background subtraction.}
\label{fig:1}
\end{figure}

\subsection{Measurement set--up}
The optical set--up is sketched in Fig. \ref{fig:1a}. It consists of a light--emitting diode (LED) as a light source, two lenses, and a digital camera. The lenses have a diameter of \SI{20}{\centi\metre}. After the measurements with the professional opera singer, the LED illumination was improved, and the recordings with the amateur singer were made with a higher light intensity. We also changed the ISO setting of the camera between these experiments.

\subsection{Data analysis}
From the raw images taken by the digital camera, we subtract the background using the algorithm KNN \citep{ZivkovicvanderHeijden2006} implemented in OpenCV. If necessary, we crop each frame to exclude facial parts. We then count the particles in either the full or cropped images. We compute a running mean of \num{100} frames to smooth the data over \SI{1}{\second} to improve the visualization. An example of an analysed raw image and the background--subtracted frame with highlighted detected particles is shown in Fig. \ref{fig:1}. Note that some particles are not detected because they are outside the cropped region (see Fig. \ref{fig:1c} for an example), though they might enter the region in the next frames and thus be detected. Dust particles are also detected, and these make up the noise---particles from musical activities generally result in prominent spikes, because many droplets are emitted at a time.

\section{Summary of initial results}
To indicate how singing, speaking, laughing, and reciting a poem compare with each other when it comes to emitting droplets, we present initial results from experiments with an amateur singer and a professional opera singer\footnote{For supplementary videos, see \citet{Dunkeretal2020data}.}. These two experiments must not be compared quantitatively, because the set--up was changed slightly between the measurements. The particle count results are relative numbers: each data point shows the ratio of detected particles to the maximum number of detected particles in the whole time series. Thus, the data span the range from \num{0} to \num{1}.

\subsection{Comparison of singing, speaking, laughing, and poetry recitation}
\begin{figure}[!t]
\centering
\includegraphics[width=0.65\textwidth]{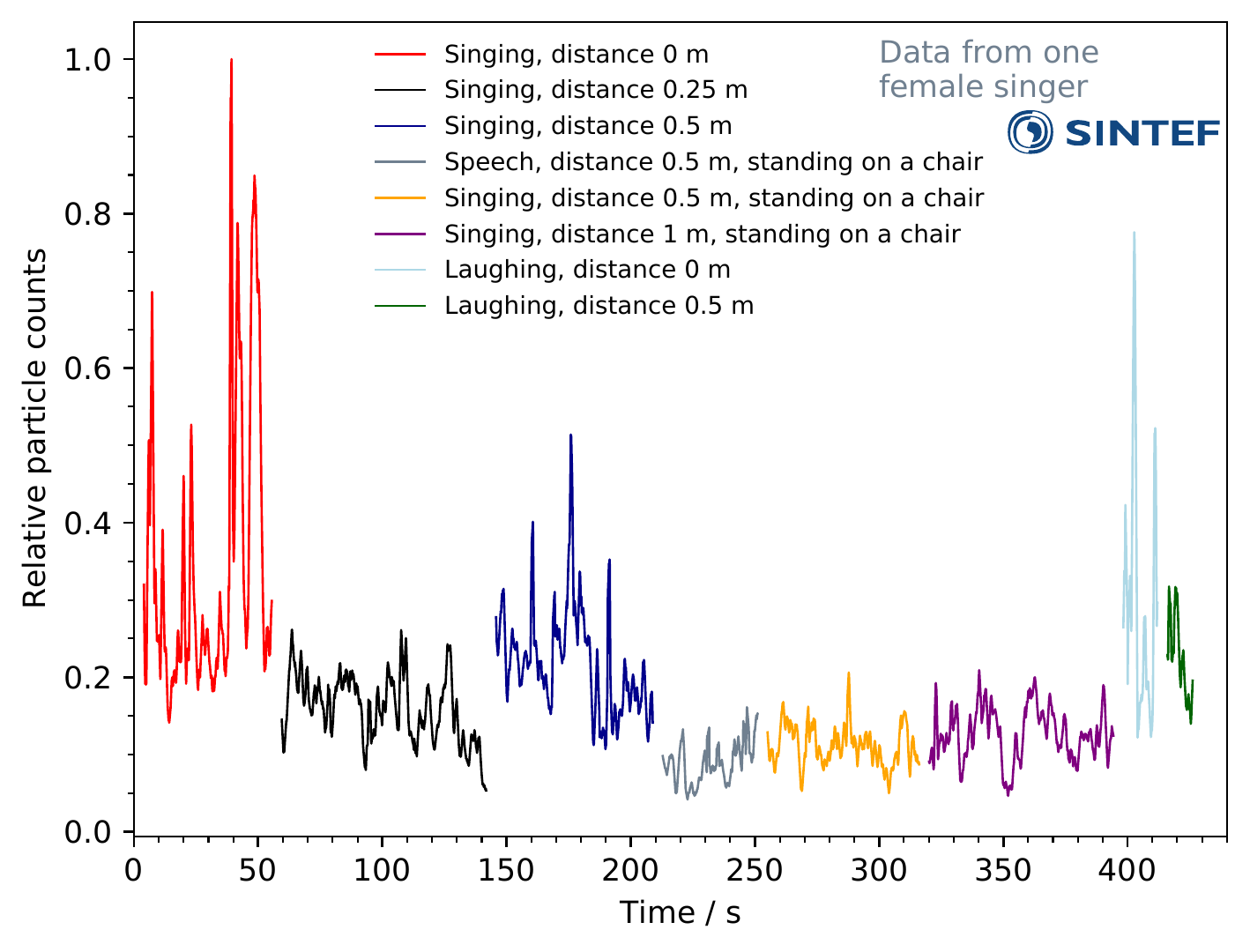}
\caption{Time series of relative number of counted particles from experiments with an amateur singer. All particle counts are relative to the maximum of the whole time series. Data are smoothed with a running mean of \SI{1}{\second}. Preliminary results, not for distribution.}
\label{fig:2}
\end{figure}

\begin{figure}[!t]
\centering
\includegraphics[width=0.65\textwidth]{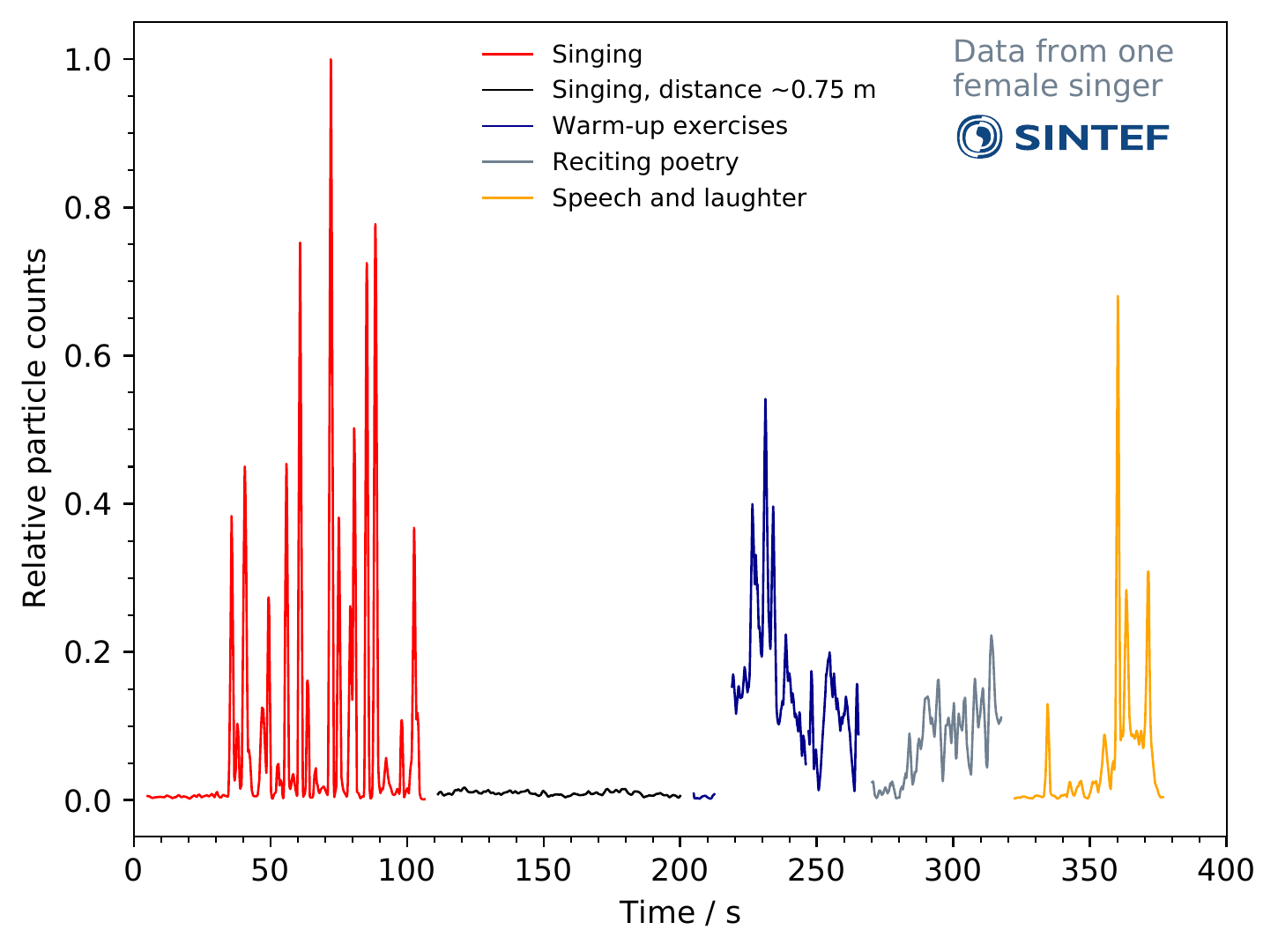}
\caption{Time series of relative number of counted particles from experiments with a professional opera singer. All particle counts are relative to the maximum of the whole time series. Periods with non--stationary reflections were removed from the time series. Data are smoothed with a running mean of \SI{1}{\second}. Preliminary results, not for distribution.}
\label{fig:3}
\end{figure}

Figure \ref{fig:2} shows counted particles during different vocal activities performed by an amateur singer. Figure \ref{fig:3} shows particles counted during different vocal activities performed by a professional opera singer. Both singers are female. Note that the two experiments were carried out in different rooms. The room in which the experiments with the opera singer were performed had considerably less dust.

In case of the amateur singer, experiments like singing, speaking, and laughing were conducted at distances of \SI{0}{\metre} (that is, the mouth is inside the measurement volume), \SI{0.25}{\metre}, \SI{0.5}{\metre}, and \SI{1}{\metre} from the measurement volume. From Fig. \ref{fig:2}, we can see that most droplets are detected when the amateur singer’s mouth is inside the measurement volume. The same exercises were repeated at the given distances. A much lower number of droplets was detected when the distance increased. To capture droplets in freefall, the singer also stood on a chair above the measurement volume and repeated the exercises. The height of the chair is about \SI{0.45}{\metre}. We did not detect a pronounced increase. When the singer laughed, however, we detected a larger number of droplets, in magnitude comparable to singing. A professional opera singer performed similar tasks in another room, which had less dust. Figure \ref{fig:3} shows the results from these experiments. The singer sang Puccini's aria ``O mio babbino caro'' twice: once with the mouth inside the measurement volume (labelled ``Singing'') and once from a distance of about \SI{0.75}{\metre} (``Singing, $\sim\SI{0.75}{\metre}$ distance''). Furthermore, the singer performed warm--up exercises with different combinations of vowels and consonants (``Warm--up exercises''), and had a conversation that included laughter (``Speech and laughter''). To resemble drama plays in a theatre, the singer recited a poem (``Reciting poetry'').

As expected, we detected most droplets when the singer's mouth was inside the measurement volume. When the singer repeated the song from a distance, we did not measure such pronounced peaks. Warm--up exercises occasionally lead to relatively as many particles as when singing the aria. Especially exercises that involvement tongue movement and rapid opening and closing of the mouth seem to generate a relatively large number of droplets. There does not seem to be a detectable difference between speaking normally and reciting a poem (Fig. \ref{fig:3}). The peaks between \SI{350}{\second} and \SI{360}{\second} in the ``Speech and laughter'' time series are due to laughter. Laughter seems to produce more droplets than conventional speaking, and at times roughly as many as singing. This is the case for both a professional and an amateur singer.

\subsection{How far can droplets reach?}
\begin{figure}[!t]
\centering
\includegraphics[width=0.75\textwidth]{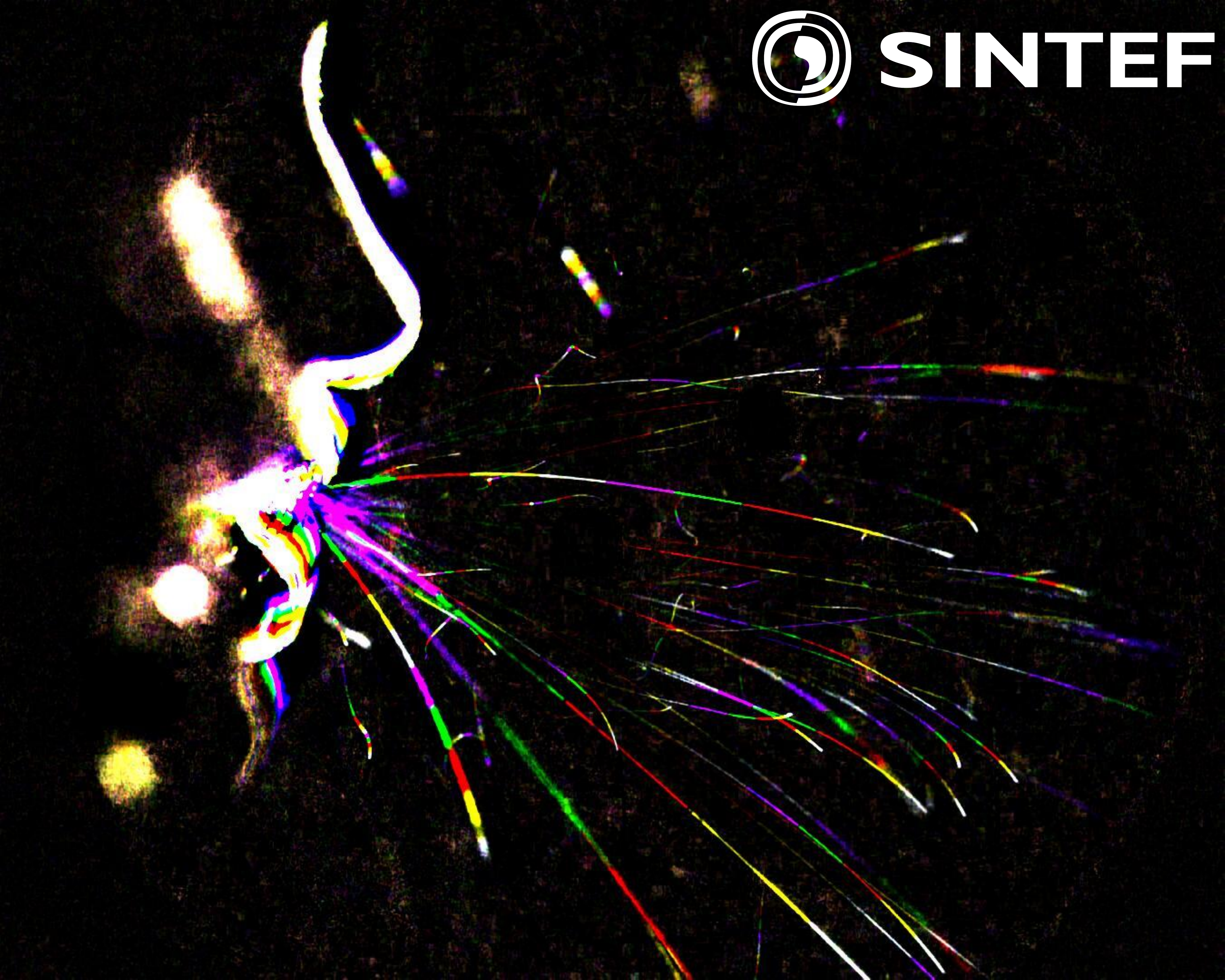}
\caption{Visualization of an extreme event of exhaled particles during singing. Note that this extreme case was chosen only for visualization---it is by no means representative (compare with Fig. \ref{fig:3}). The dimensions of the image are approximately \SI{20}{\centi\metre} $\times$ \SI{20}{\centi\metre}. Droplets that seemingly reach the end of their trajectory on the right side simply leave the measurement volume. The image is a composite of ten consecutive frames, each with an exposure time of \SI{0.01}{\second}. We used a different background subtraction than in the other images shown here.}
\label{fig:4}
\end{figure}

Figure \ref{fig:4} illustrates an extreme example of droplets produced while singing. Note that most droplets have a downward trajectory after leaving the mouth. From this very limited dataset, we can see that the number of detectable droplets decreases with distance from the measurement volume (see Fig. \ref{fig:2}). This is as expected. Overall, the data collected suggest that the relative number of droplets when the singer is \SI{0.5}{\metre} from the measurement volume is less than \SI{50}{\percent} compared to when the singer's face is inside the measurement volume.

To determine how far droplets can reach, we performed several experiments with the amateur singer. These experiments involved warm--up exercises and laughter at various distances (including standing on a podium) beside the measurement volume. Results are shown in Fig. \ref{fig:2}.

There is no clear difference between an amateur and a professional singer in the cases considered here. Even at a distance of \SI{0.25}{\metre} from the measurement volume, the number of detected droplets is far lower compared to the case of the face being immediately inside the measurement volume. It seems as if there are somewhat more droplets when singing at \SI{0.5}{\metre} compared to \SI{0.25}{\metre} in the case of the amateur singer: this could be due to more dust, somewhat more droplets from singing due to a wetter mouth, or a combination of both. We detected approximately as many particles when the singer stood on a platform at distances of \SI{0.5}{\metre} and \SI{1}{\metre}. When standing on the platform at \SI{0.5}{\metre} m from the measurement volume, we would have expected an increase in the number of detected droplets, because the droplets of the same size and energy could travel farther before hitting the ground due to gravity. This discussion illustrates the necessity of more measurements under different circumstances.

By assuming that droplets follow a ballistic trajectory, we can estimate the distance they travel before hitting the ground. The distance varies with initial height. Given the same exit velocity and size of a droplet, this droplet will reach farther when emitted by a taller person or by a person standing on a podium.

We have detected droplets with diameters of \SI{50}{\micro\metre} and above. Once ejected, the droplets start to slow down to terminal velocity because of friction with air molecules (compare Figs. \ref{fig:4} and \ref{fig:5}). This terminal velocity will depend on droplet size. Preliminary calculations, considering the initial speed and estimated terminal velocity of the droplets, indicate that most of the droplets will have hit the ground before travelling a horizontal distance of \SI{1}{\metre}.

\subsection{Playing musical instruments}

\begin{figure}[!t]
\centering
\includegraphics[width=0.75\textwidth]{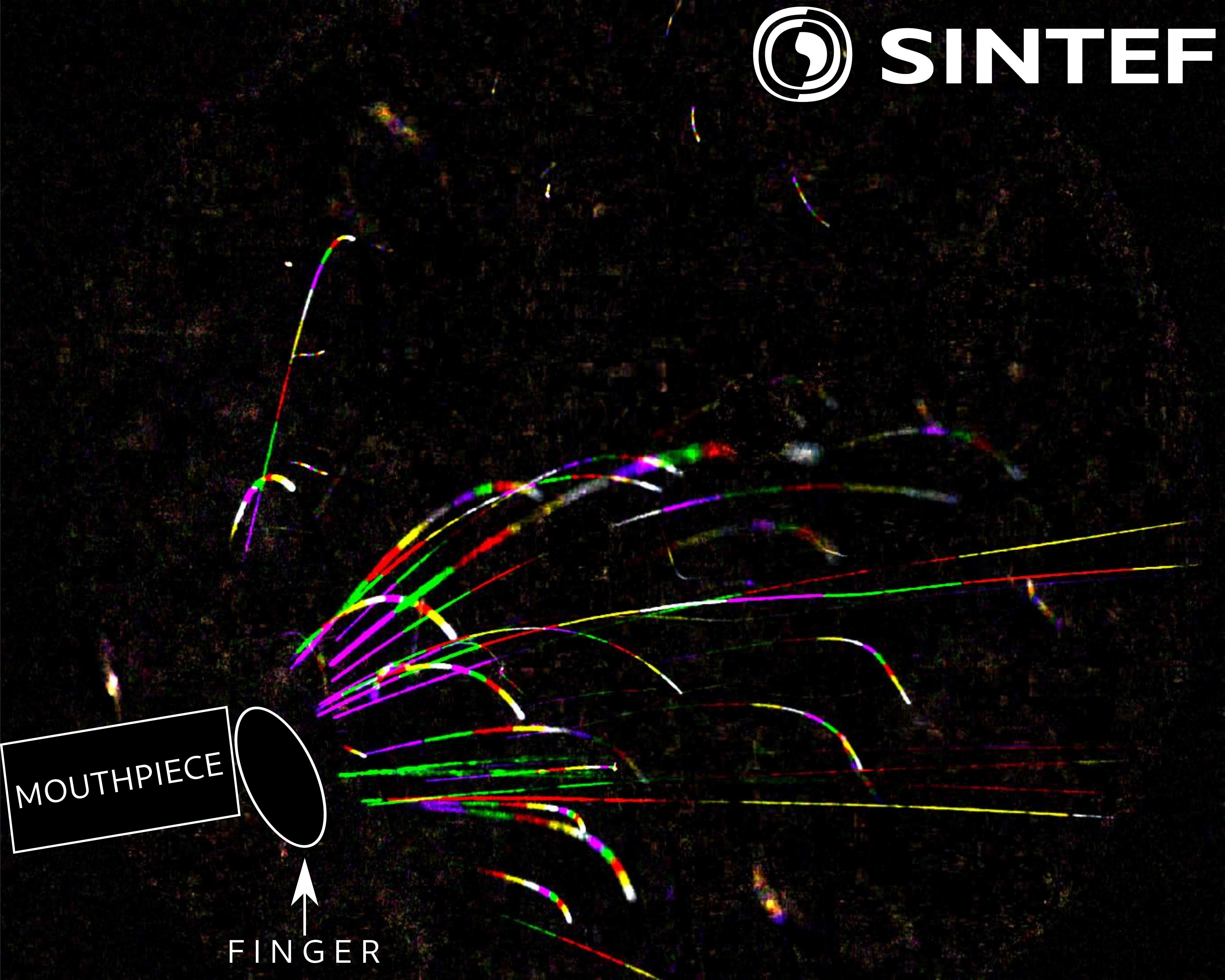}
\caption{Examples of droplets ejected from a tuba's mouthpiece when the opening was partially blocked by a finger. The mouthpiece was held horizontally. The image is a composite of ten consecutive frames with an exposure time of \SI{0.01}{\second} each. The dimensions of the image are approximately \SI{20}{\centi\metre} by \SI{20}{\centi\metre}.}
\label{fig:5}
\end{figure}

We have recorded videos of musicians playing the flute, the clarinet, and the tuba. In the latter case, the tubist also performed a warm--up exercises using only the mouthpiece.

We detected only a few droplets that left the opening of the clarinet. The clarinet pointed downwards at an angle of approximately 45 degrees. These droplets thus followed a downward trajectory, hitting the ground after a short distance.

Likewise, we did not detect many droplets from the flute. This is true for both the opening and the mouthpiece region of the instrument.

We did not detect droplets from the tuba itself. The tuba player warmed up by using only the mouthpiece, that is, the mouthpiece was detached from the tuba. Figure \ref{fig:5} shows an extreme event with many droplets exiting the tuba's mouthpiece. The mouthpiece was held horizontally, with a finger partially blocking the opening. This is a typical warm--up exercise for musicians. Figure \ref{fig:5} shows that droplets with different trajectories leave the mouthpiece, and that many droplets are slowed down after a short distance and then fall down. Some particles are rather fast and thus follow trajectories that have only a small downward component at first. When the mouthpiece was held without a finger on the opening, we did detect some droplets, but not as many and with shorter trajectories.

\section{Conclusions}
First, we must emphasize that this memo contains initial results from a very small sample. The results should be interpreted accordingly. Furthermore, no conclusions can be drawn from our results about possible risks of infection with a disease. All experiments were conducted without a facemask.

We have built a portable measurement set--up to detect and track droplets. First experiments with one professional opera singer and one amateur singer showed that:
\begin{itemize}
\item Singing produces more droplets relative to speaking or poetry recitation, with certain sounds producing more droplets than others.
\item Laughing can lead to more droplets than speaking and is comparable in magnitude to singing.
\item We detected considerably fewer droplets from the same warm--up exercise when distance was increased to \SI{0.25}{\metre} and \SI{0.5}{\metre}. However, because the measured droplets follow a ballistic trajectory, they can reach a distance of up to one meter before hitting the ground---depending on the singer's height and the droplet's size and speed.
\end{itemize}

We have also performed experiments with a tuba, a flute, and a clarinet:
\begin{itemize}
\item We did not detect droplets from a tuba. From the mouthpiece alone, however, droplets can be ejected during warm--up exercises. If the opening is partially blocked (e.g., by a finger) and the mouthpiece is not pointed downwards, ejected droplets can travel a considerable distance before hitting the ground, depending on their size and speed. Partial blocking leads to a build--up of pressure inside the mouthpiece.
\item We did not detect a considerable number of droplets from a flute or a clarinet. The clarinet usually points downwards at some angle, such that ejected droplets follow a downward trajectory.
\end{itemize}
These results are from initial experiments. The actual conditions (whether the singer has recently drunk water; different voice or volume; et cetera) probably influence the number of droplets generated during the vocal activities.

\begin{acknowledgements}
\footnotesize
This project is partially funded by the Norwegian Choir Association (Norges Korforbund). The Norwegian Choir
Association had no role in the design of the experiments, data analysis and interpretation, and writing this document. It is a pleasure to acknowledge the fruitful discussions with Ernst--Kristian R\o dland of the Norwegian Public Health Administration (Folkehelseinstituttet). Co--sponsors of this project are:
\begin{itemize}
\item Norges Korforbund \slash Norwegian Choir Association
\item Musikkens studieforbund \slash The Adult Education Association of Music in Norway
\item Krafttak for sang \slash Singing Norway
\item Norsk musikkr\aa d \slash The Norwegian Music Council
\item Norsk teater-- og orkesterforening \slash Association of Norwegian Theatres and Orchestras
\item Den Norske Opera \& Ballett \slash The Norwegian Opera and Ballet
\item Norges teknisk--naturvitenskapelige universitet \slash Norwegian University of Science and Technology
\item Creo -- forbundet for kunst og kultur \slash Creo -- the union for arts and culture
\item Den norske kirke \slash Church of Norway
\item Norges Musikkorps Forbund \slash Norwegian Band Federation
\item Arktisk Filharmoni \slash Arctic Philharmonic
\item Frilynt Norge
\item Kulturalliansen
\end{itemize}

\end{acknowledgements}

\normalsize
\bibliography{droplets.bib}

\end{document}